\documentclass[letter]{IEEEtran}
\usepackage{cmap}
\usepackage{amsmath}
\usepackage{amsfonts}
\usepackage{subfigure}
\usepackage{graphics,graphicx,picture}
\usepackage{mathrsfs}
\usepackage{cases}
\usepackage{bm, comment,epstopdf,color,amssymb}
\usepackage{cite}
\usepackage{algorithm}
\usepackage{algorithmic}
\usepackage{tabularx}

\begin{document}

\title{Joint Beamforming Design in Multi-Cluster MISO NOMA Intelligent Reflecting Surface-Aided Downlink Communication Networks}

\author{Yiqing Li, Miao Jiang, Qi Zhang, \emph{Member}, \emph{IEEE}, and Jiayin Qin
	
\thanks{Y. Li, M. Jiang, Q. Zhang, and J. Qin are with the School of Electronics and Information Technology, Sun Yat-sen University, Guangzhou 510006, China (e-mail: liyiq5@mail2.sysu.edu.cn; jmiao@mail2.sysu.edu.cn; zhqi26@mail.sysu.edu.cn; issqjy@mail.sysu.edu.cn).}

}

\markboth{}
{Li \MakeLowercase{\textit{et al.}}: Joint Beamforming Design in Multi-Cluster MISO NOMA IRS-Aided Downlink Communication Networks}

\maketitle
\begin{abstract}
Considering intelligent reflecting surface (IRS), we study a multi-cluster multiple-input-single-output (MISO) non-orthogonal multiple access (NOMA) downlink communication network. In the network, an IRS assists the communication from the base station (BS) to all users by passive beamforming. Our goal is to minimize the total transmit power by jointly optimizing the transmit beamforming vectors at the BS and the reflection coefficient vector at the IRS. Because of the restrictions on the IRS reflection amplitudes and phase shifts, the formulated quadratically constrained quadratic problem is highly non-convex. For the aforementioned problem, the conventional semidefinite programming (SDP) based algorithm has prohibitively high computational complexity and deteriorating performance. Here, we propose an effective second-order cone programming (SOCP)-alternating direction method of multipliers (ADMM) based algorithm to obtain the locally optimal solution. To reduce the computational complexity, we also propose a low-complexity zero-forcing (ZF) based suboptimal algorithm. It is shown through simulation results that our proposed SOCP-ADMM based algorithm achieves significant performance gain over the conventional SDP based algorithm. Furthermore, when the number of passive reflection elements is relatively high, our proposed ZF-based suboptimal algorithm also outperforms the SDP based algorithm.
\end{abstract}

\begin{IEEEkeywords}
Alternating direction method of multipliers (ADMM), intelligent reflecting surface (IRS), multiple-input-single-output (MISO), non-orthogonal multiple access (NOMA), zero-forcing (ZF).
\end{IEEEkeywords}
\IEEEpeerreviewmaketitle

\section{Introduction}

Recently, intelligent reflecting surface (IRS) had been envisioned as a cost-effective and green solution to significantly increase spectrum and energy efficiencies \cite{JZhao}. An IRS, equipped with reconfigurable passive reflection elements, can reconfigure the propagation environment by reflecting the incoming signals in a programmable manner with low power consumption \cite{LLi,FLiu}, and nearly no additional thermal noise is included during the reflection procedure.
	
The existing literature has studied the problems related to IRS in various areas, e.g., channel estimation, capacity analyses, power optimization, secure transmission, deep-learning based design, etc. \cite{QUANadeem,DMishra,SHu,HGuo,CHuang,MCui,ATaha}. In \cite{QUANadeem}, a minimum mean squared error (MMSE) based channel estimation protocol was proposed in the  IRS-assisted multi-user multiple-input-multiple-output (MIMO) communication system. In \cite{DMishra}, using IRS to support wireless energy transfer from a multiple-antenna power beacon (PB) to a single-antenna user was proposed. In \cite{SHu}, it was shown that the capacity achieved per square meter surface-area is linearly proportional to the average transmit power when the surface-area of the IRS is sufficiently large. By jointly optimizing the active beamforming at the base station (BS) and the passive beamforming at the IRS, weighted sum-rate maximization in IRS-aided multiple-input-single-output (MISO) was investigated in \cite{HGuo}. In \cite{CHuang}, alternating maximization was proposed for transmit power allocation at the BS and the reflection coefficients at the IRS in IRS-aided downlink multi-user MISO system. In \cite{MCui}, the secrecy rate maximization problem was considered where a multi-antenna BS communicates with a single-antenna user in the presence of a single-antenna eavesdropper. In \cite{ATaha}, the deep learning-based IRS channel estimation was studied by treating the wireless propagation as a deep neural network, the IRS units as neurons, and cross-interactions between units as links.
	
To promote IRS to be integrated into future wireless communications, multiple access techniques are essential. Currently, non-orthogonal multiple access (NOMA) is regarded as a promising wireless radio access technology since it can support multiple users in the same resource block such as time, frequency, and code \cite{MFHanif,YLi,YLi2,STimotheou,ZDing00,WLiang,Zding01}. In NOMA communication networks, superimposed signals are transmitted from the transmitter and successive interference cancellation (SIC) is applied at the receiver for interference cancellation. The earlier works show that NOMA can achieve considerable performance gains over orthogonal multiple access (OMA) in terms of sum rate, secrecy rate, user fairness, and outage probability, etc. \cite{MFHanif,YLi,YLi2,STimotheou,ZDing00,WLiang,Zding01}.
	
Inspired by the aforementioned potential benefits of IRS and NOMA, it is interesting to investigate the promising applications of NOMA technique in the IRS enhanced wireless network to further improve the system performance \cite{YangG,MFu}. In \cite{YangG}, Yang \emph{et al.} considered an IRS-assisted downlink NOMA communication system where a single-antenna BS transmits superposed signals to multiple single-antenna users via the NOMA protocol. In the proposed scheme in \cite{YangG}, semidefinite relaxation (SDR) technique followed by Gaussian randomization was applied during the phase shift optimization. In \cite{MFu}, Fu. \emph{et al.} investigated total transmit power minimization problem for the NOMA downlink transmission where an IRS was deployed to assist the transmission from a multi-antenna BS to multiple single-antenna users.
	
Note that in \cite{MFu}, a single cluster was considered. In many practical scenarios, users are encouraged to group into small-size clusters to lower the  decoding complexity at each user \cite{ZDing00,ZDing02}. In this paper, we focus on a multi-cluster MISO NOMA IRS-aided downlink communication network. In the network,
a multi-antenna BS transmits the superimposed signals to multiple clusters of single-antenna users. Each cluster includes one cell-edge user and one central user. An IRS with multiple low cost passive elements is deployed to steer the incident signals by passive beamforming and assist the communication from the BS to all users. Different cases which restricts the IRS reflection amplitudes and phase shifts are considered.

We aim to minimize the total transmit power at the BS by jointly optimizing the transmit beamforming vectors at the BS and the reflection coefficient vector at the IRS. The formulated quadratically constrained quadratic problem (QCQP) is highly non-convex because of the restrictions on the IRS reflection amplitudes and phase shifts. A common way of solving the non-convex QCQP is to convert it into a convex semidefinite programming (SDP) via SDR \cite{QWu}. However, the computational complexity of SDP based algorithm is prohibitively high because the number of involved variables is quadratic in the number of reflection elements. Furthermore, the probability to obtain a rank-one optimum solution to the problem after SDR is extremely small. In this paper, we propose an effective second-order cone programming (SOCP)-alternating direction method of multipliers (ADMM) based algorithm to obtain the locally optimal solution. In the proposed scheme, we introduce auxiliary variables to transform the achievable rate constraints into quadratic constraints. They are further transformed into second-order cone constraints. Using the ADMM method \cite{SBoyd}, the formulated QCQP is decomposed into multiple SOCPs. To reduce the computational complexity, we also propose a low-complexity zero-forcing (ZF) based suboptimal algorithm.
	
\emph{Notations}: Boldface lowercase and uppercase letters denote vectors and matrices, respectively. The transpose, conjugate, conjugate transpose, and trace of the matrix $\mathbf{A}$ are denoted as $\mathbf{A}^T$, $\mathbf{A}^*$, $\mathbf{A}^{H}$, and $\mbox{tr}(\mathbf{A})$, respectively. By $\mathbf{A}\succeq\mathbf{0}$, we mean that $\mathbf{A}$ is positive semidefinite. $\mathcal{CN}(0,\sigma^2)$ denotes the distribution of a circularly symmetric complex Gaussian random variable with zero mean and variance $\sigma^2$. $\left\|\mathbf{a}\right\|_2$ denotes the Euclidean norm of a complex vector $\mathbf{a}$. $\mbox{Round}(a)$ denotes the round-to-integer operation on a real value $a$. $\angle a$ denotes the angle of a complex value $a$. The Hadamard division between two vectors is denoted by $\oslash$. The operation $\mbox{Abs}(\mathbf{a})$ constructs a vector by extracting the magnitudes of all the elements of vector $\mathbf{a}$. $\mbox{Bdiag}[\mathbf{A}_1,\cdots,\mathbf{A}_b]$ denotes a block diagonal matrix where the diagonal blocks are $\mathbf{A}_1$, $\cdots$, $\mathbf{A}_b$.

\section{System Model and Problem Formulation}

Consider a multi-cluster MISO NOMA IRS-aided downlink communication network, where the BS with $N$ antennas serves $2K$ single-antenna users with an IRS. For the sake of improving the spectrum efficiency and reducing the system load, the users are grouped into $K$ clusters where the $k$-th cluster, $k\in\mathcal{K}=\{1, \cdots, K\}$, includes one cell-edge user $U_{k,e}$ and one central user $U_{k,c}$. In each cluster, the NOMA protocol is applied during transmission. We assume that specific user pairing strategy during the NOMA transmission \cite{ZDing00,WLiang,Zding01} is employed. Our main focus is on joint beamforming design after user pairing. To enhance the transmission performance, a large IRS with $M$ passive and low-cost phase shifters is installed on a surrounding building to assist the BS in communicating with the users. The phase shifts of the IRS are programmable and configurable via an IRS controller. Therefore, each user receives the superposed signals from both the BS-user (direct) link and IRS-user (reflect) link.

For the considered MISO NOMA IRS-aided downlink communication network, the complex baseband signal transmitted from the BS is formulated as
\begin{align}
\mathbf{x} = \sum_{k = 1}^{K}\left( \mathbf{w}_{k,c} s_{k,c} + \mathbf{w}_{k,e}s_{k,e}\right)
\end{align}
where $s_{k,c}$ and $s_{k,e}$ are defined as the transmitted symbols for the central user and cell-edge user in the $k$-th cluster, respectively, with $\mathbb{E}(s_{k,c}^Hs_{k,c}) = \mathbb{E}(s_{k,e}^Hs_{k,e}) = 1$; $\mathbf{w}_{k,c}\in \mathbb{C}^{N \times 1}$ and $\mathbf{w}_{k,e} \in \mathbb{C}^{N \times 1}$ denote the beamforming vectors intended for the central user and cell-edge user in the $k$-th cluster, respectively.

The complex-valued received signal at $U_{k,i}$, $k \in \mathcal{K}$, $i\in \{c,e\}$, can be mathematically expressed as
\begin{align}
y_{k,i} = \left(\mathbf{g}_{k,i}^H\bm{\Phi}\mathbf{H} + \mathbf{h}_{k,i}^H\right)\mathbf{x} + n_{k,i}
\end{align}
where $\mathbf{H} \in \mathbb{C}^{M \times N}$, $\mathbf{g}_{k,i} \in \mathbb{C}^{M \times 1}$ and $\mathbf{h}_{k,i} \in
\mathbb{C}^{N \times 1}$ denote the baseband equivalent channels from the BS to the IRS, from the IRS to $U_{k,i}$, and from the BS to $U_{k,i}$, respectively;
$\bm{\Phi} = \mbox{diag}\left(\bm{\phi}\right)$ denotes the diagonal reflection coefficients matrix of the IRS with
\begin{equation}\label{q3}
\bm{\phi} = \left[\phi_1, \cdots, \phi_M\right]^T;
\end{equation}
$n_{k,i} \sim \mathcal{CN}(0, \sigma_{k,i}^2)$ denotes the additive Gaussian noise at $U_{k,i}$. Without loss of generality, we assume $\sigma_{k,i}^2 = \sigma^2$ for $k\in\mathcal{K}$ and $i\in\{c,e\}$ throughout the paper. In \eqref{q3}, we have \cite{LiangYC}
\begin{equation}
\phi_m= \alpha_me^{j \theta_m}
\end{equation}
for $m \in\mathcal{M}=\{1, \cdots, M\}$, where $\theta_m \in \left[0,2\pi\right]$ and $\alpha_m$ are the phase shift and amplitude reflection coefficient on the incident signal, respectively.

Accordingly, the signal-to-interference-and-noise-ratio (SINR) to decode $s_{k,e}$ at $U_{k,e}$ can be expressed as
\begin{align}\label{q5}
\gamma_{k,e}=\frac{|\hat{\mathbf{h}}_{k,e}^H \mathbf{w}_{k,e}|^2}{\sigma^2 + |\hat{\mathbf{h}}_{k,e}^H\mathbf{w}_{k,c}|^2 +\zeta_{k,e}}
\end{align}
where $\hat{\mathbf{h}}_{k,i}^H = \mathbf{g}_{k,i}^H \bm{\Phi}\mathbf{H} + \mathbf{h}_{k,i}^H$ and
\begin{equation}
\zeta_{k,i}=\sum_{j=1, j\neq k}^{K}\left(|\hat{\mathbf{h}}_{k,i}^H\mathbf{w}_{j,c}|^2 + |\hat{\mathbf{h}}_{k,i}^H\mathbf{w}_{j,e}|^2\right).
\end{equation}
Similarly, the SINR for $U_{k,c}$ to decode $s_{k,e}$ is given by
\begin{align}\label{q7}
\gamma_{k,c,e} = \frac{|\hat{\mathbf{h}}_{k,c}^H \mathbf{w}_{k,e}|^2}{\sigma^2 + |\hat{\mathbf{h}}_{k,c}^H\mathbf{w}_{k,c}|^2 +\zeta_{k,c}}.
\end{align}

Following the NOMA protocol, if $U_{k,c}$ can decode the symbol $s_{k,e}$ successfully before decoding its own symbol $s_{k,c}$, SIC can be employed. After perfect SIC, the SINR to decode the symbol to itself, $s_{k,c}$, at $U_{k,c}$ can be expressed as
\begin{align}\label{q8}
\gamma_{k,c}=\frac{|\hat{\mathbf{h}}_{k,c}^H \mathbf{w}_{k,c}|^2}{\sigma^2 +\zeta_{k,c}}
\end{align}
where the inter-user interference in the same cluster has been removed.

In this paper, we aim to minimize the total transmit power at the BS by jointly optimizing the transmit beamforming vectors $\{\mathbf{w}_{k,i}\}$ at the BS and reflection coefficient vector $\bm{\phi}$ at the IRS, subject to the transmission rate requirements at $2K$ users. The corresponding optimization problem is formulated as
\begin{subequations}\label{q9}
\begin{align}
\label{q9a}\min_{\bm{\phi}, \{\mathbf{w}_{k,i}\}}\ & \sum_{k = 1}^{K} \left( \left\|\mathbf{w}_{k,c}\right\|^2+ \left\|\mathbf{w}_{k,e}\right\|^2\right) \\
\label{q9b} \mbox{s.t.}\ \ & \log_2\left(1 + \gamma_{k,c} \right) \geq r_{k,c}, \ \forall\ k\in\mathcal{K},\\
\label{q9c} & \log_2\left(1 + \min\left(\gamma_{k,e}, \gamma_{k,c,e}\right)\right) \geq r_{k,e}, \ \forall\ k\in\mathcal{K},\\
\label{q9d} &  \bm{\phi} \in \mathcal{F}
\end{align}
\end{subequations}
where $r_{k,c}$ and $r_{k,e}$ denote the target transmission rates of $U_{k,c}$ and $U_{k,e}$ in the $k$-th cluster, respectively; $\mathcal{F}$ denotes the feasible set of the reflection coefficient vector. Typically, there are three different assumptions for the reflection coefficient vector $\bm{\phi}$ and feasible set $\mathcal{F}$ can be listed as follows \cite{LiangYC}:
\begin{align}\label{q10}
\mathcal{F} =
\begin{cases}
\left\{\bm{\phi} \big| |\phi_m|^2 \in \left[0,1\right] \right\}; & \text{Case I},\\
\left\{\bm{\phi} \big| |\phi_m|^2 = 1\right\}; &  \text{Case II}, \\
\left\{\bm{\phi} \big| |\phi_m|^2 = 1;\ \theta_m \in \mathcal{A}\right\}; & \text{Case III}.
\end{cases}
\end{align}
In \eqref{q10}, Case I denotes the case where both the amplitude and phase shift are continuous variables, Case II denotes the case where the amplitude is fixed and phase-shift is continuous, and Case III denotes the case where the amplitude is fixed and phase-shift is discrete,
\begin{equation}
\mathcal{A} = \left\{0,\ 2\pi/L,\ \cdots,\  2\pi\left(L- 1\right)/L \right\}
\end{equation}
in which $L$ denotes the number of quantized reflection coefficient values of the reflective-radio elements on the IRS. Due to the coupled variables in the constraints, the optimization problem \eqref{q9} is non-convex.

\section{SOCP-ADMM Based Algorithm}
In this section, we propose an SOCP-ADMM based algorithm to find the locally optimal solution to the optimization problem \eqref{q9}. To continue, we need following lemma \cite{YuW1,YuW2}.

\textit{Lemma 1:} For $\mathbf{A} \in \mathbb{C}^{a \times b}$, $\mathbf{B} \in \mathbb{C}^{b \times b}$ and $\mathbf{B} = \mathbf{B}^H$, we have
\begin{align} \label{bq1}
\mbox{tr}\left(\mathbf{A}^H\mathbf{B}^{-1}\mathbf{A}\right) = \underset{\mathbf{Y}}{\max} \ \mbox{tr}\left(\mathbf{Y}^H\mathbf{A} + \mathbf{A}^H\mathbf{Y} - \mathbf{Y}^H\mathbf{B}\mathbf{Y}\right)
\end{align}
where $\mathbf{Y} \in \mathbb{C}^{a \times b}$.

\textit{Proof}: Let $f\left(\mathbf{Y}\right) = \mbox{tr}\left(\mathbf{Y}^H\mathbf{A} + \mathbf{A}^H\mathbf{Y} - \mathbf{Y}^H\mathbf{B}\mathbf{Y}\right)$. Since $f\left(\mathbf{Y}\right)$ is concave in terms of $\mathbf{Y}$, the optimal value of $f\left(\mathbf{Y}\right)$ can be obtained by letting $\frac{\partial f\left(\mathbf{Y}\right)}{\partial \mathbf{Y}} = \mathbf{0}$. Therefore, the optimal $\mathbf{Y}^o$ is $\mathbf{B}^{-1}\mathbf{A}$. By replacing the obtained $\mathbf{Y}^o$ into $f\left(\mathbf{Y}\right)$, \eqref{bq1} can be obtained.
$\hfill\blacksquare$

Using Lemma 1 and introducing auxiliary variables $u_{k,1} \in
\mathbb{C}^{1 \times 1}$, $k \in \mathcal{K}$, the constraints in \eqref{q9b} can be transformed into
\begin{align}
\label{bq2}
&\max_{u_{k,1}}\ 2\mbox{Re}\left\{u_{k,1}^H\hat{\mathbf{h}}_{k,c}^H\mathbf{w}_{k,c} \right\} - u_{k,1}^H\beta_{k,1}u_{k,1} \geq \tau_{k,c},\ \forall\ k\in\mathcal{K}
\end{align}
where $\beta_{k,1}=\sigma^2+\zeta_{k,c}$ and $\tau_{k,c}=2^{r_{k,c}}-1$. Similarly, introducing auxiliary variables $u_{k,2} \in
\mathbb{C}^{1 \times 1}$ and $u_{k,3} \in
\mathbb{C}^{1 \times 1}$, $k \in \mathcal{K}$, the constraints in \eqref{q9c} can be transformed into
\begin{align}
\label{bq3}
&\max_{u_{k,2}}\ 2\mbox{Re}\left\{u_{k,2}^H\hat{\mathbf{h}}_{k,e}^H\mathbf{w}_{k,e} \right\} - u_{k,2}^H\beta_{k,2}u_{k,2} \geq \tau_{k,e},\ \forall\ k\in\mathcal{K},\\
\label{bq4}
&\max_{u_{k,3}}\ 2\mbox{Re}\left\{u_{k,3}^H\hat{\mathbf{h}}_{k,c}^H\mathbf{w}_{k,e} \right\} - u_{k,3}^H\beta_{k,3}u_{k,3} \geq \tau_{k,e},\ \forall\ k\in\mathcal{K},
\end{align}
where $\tau_{k,e}=2^{r_{k,e}}-1$,
\begin{align}
\beta_{k,2}=\sigma^2+|\hat{\mathbf{h}}_{k,e}^H\mathbf{w}_{k,c}|^2+\zeta_{k,e},\\
\beta_{k,3}=\sigma^2+|\hat{\mathbf{h}}_{k,c}^H\mathbf{w}_{k,c}|^2+\zeta_{k,c}.
\end{align}
Therefore, the original problem in \eqref{q9} can be recast as
\begin{align}\label{bq7}
\min_{\scriptsize\begin{array}{c}
\bm{\phi}, \{\mathbf{w}_{k,i}\}, \\ \{u_{k,1},u_{k,2},u_{k,3}\}
\end{array}} \ & \sum_{k = 1}^{K} \left( \left\|\mathbf{w}_{k,c}\right\|^2+ \left\|\mathbf{w}_{k,e}\right\|^2\right) \nonumber\\
\mbox{s.t.}\quad \quad \quad &
\eqref{bq2},\ \eqref{bq3},\ \eqref{bq4},\ \bm{\phi} \in \mathcal{F}.
\end{align}
Since problem \eqref{bq7} is non-convex, we propose to use the ADMM method to decompose problem \eqref{bq7} into several sub-problems. To begin with, an auxiliary variable $\bm{\varphi}$ is introduced such that
\begin{align}
\bm{\phi}=\bm{\varphi}
\end{align}
where $\bm{\varphi} \in \mathbb{C}^{M \times 1}$ is a copy of the original reflection coefficient vector $\bm{\phi}$.

Define the feasible region of the constraints \eqref{bq2}, \eqref{bq3}, and \eqref{bq4} as $\mathcal{B}$, whose indicator function is given as
\begin{align}
\mathbb{I}_{\mathcal{B}}\left(\bm{\phi}, \mathbf{w}, \mathbf{u}\right) = \begin{cases}
\|\mathbf{w}\|^2; & \mbox{if} \ \{\bm{\phi}, \mathbf{w}, \mathbf{u}\} \in \mathcal{B},  \\
\infty;  & \mbox{otherwise},
\end{cases}
\end{align}
where $\mathbf{w} = \left[\mathbf{w}_{1,c}^T, \mathbf{w}_{1,e}^T, \cdots, \mathbf{w}_{K,c}^T, \mathbf{w}_{K,e}^T \right]^T$ and $\mathbf{u} = \left[u_{1,1}, \cdots, u_{K,1}, u_{1,2}, \cdots, u_{K,2}, u_{1,3}, \cdots, u_{K,3}\right]^T$.

Similarly, we define the indicator function of the feasible region of $\mathcal{F}$ as
\begin{align}
\mathbb{I}_{\mathcal{F}}\left(\bm{\varphi}\right) =
\begin{cases}
0; & \mbox{if}\ \bm{\varphi}\in \mathcal{F}, \\
\infty; & \mbox{otherwise}.
\end{cases}
\end{align}
Thus, we obtain the equivalent ADMM reformulation of problem \eqref{bq7} as follows
\begin{align}
\label{bq11}
\min_{\bm{\phi}, \mathbf{w}, \mathbf{u}, \bm{\varphi}} \ & \mathbb{I}_{\mathcal{B}}\left(\bm{\phi}, \mathbf{w}, \mathbf{u}\right) + \mathbb{I}_{\mathcal{F}}\left(\bm{\varphi}\right) \nonumber\\
\mbox{s.t.}\quad & \bm{\phi}=\bm{\varphi}.
\end{align}

The augmented Lagrangian of problem \eqref{bq11} can be formulated as
\begin{align} \label{bq12}
\mathcal{L}_{\xi}=\mathbb{I}_{\mathcal{B}}\left(\bm{\phi}, \mathbf{w}, \mathbf{u}\right) + \mathbb{I}_{\mathcal{F}}\left(\bm{\varphi}\right) + \frac{\xi}{2}\left\|\bm{\phi}-\bm{\varphi}+\bm{\lambda} \right\|_2^2
\end{align}
where $\xi>0$ is the penalty parameter and $\bm{\lambda}=\left[\lambda_1, \cdots, \lambda_M\right]^T$ denotes the scaled dual variable vector for the constraint $\bm{\phi}=\bm{\varphi}$. Applying the ADMM method, $\mathcal{L}_{\xi}$ can be minimized by updating $\bm{\phi}$, $\mathbf{w}$, $\mathbf{u}$, $\bm{\varphi}$, and $\bm{\lambda}$, alternatively.

In the $v$-th iteration, given $\bm{\phi}^{\left(v\right)}$, $\mathbf{w}^{\left(v\right)}$, $\mathbf{u}^{\left(v\right)}$, $\bm{\varphi}^{\left(v\right)}$, and $\bm{\lambda}^{\left(v\right)}$, the details of updating each variable are as follows.

\textit{1) Update $\bm{\phi}$}: The sub-problem for updating variable $\bm{\phi}$ is expressed as
\begin{align} \label{cq1}
\bm{\phi}^{\left(v+1\right)}= \arg\underset{\bm{\phi}}{\min}\ \mathcal{L}_{\xi}.
\end{align}
Problem \eqref{cq1} is equivalent to
\begin{align}
\label{cq2}
\min_{\bm{\phi}} \ & \left\|\bm{\phi} - \bm{\varphi}^{\left(v\right)} + \bm{\lambda}^{\left(v\right)} \right\|_2^2 \nonumber\\
\mbox{s.t.}\ & \eqref{bq2},\ \eqref{bq3},\ \eqref{bq4}.
\end{align}
By letting $\mathbf{G}_{k,i} = \mbox{diag}(\mathbf{g}_{k,i}^H)$, we rewrite $\hat{\mathbf{h}}_{k,i}^H$ as
\begin{equation}
\hat{\mathbf{h}}_{k,i}^H=\bm{\phi}^T\mathbf{G}_{k,i}\mathbf{H} + \mathbf{h}_{k,i}^H
\end{equation}
for $k\in\mathcal{K}$, $i\in\{c,e\}$. Substituting $\mathbf{w}^{\left(v\right)}$, $\mathbf{u}^{\left(v\right)}$, $\bm{\varphi}^{\left(v\right)}$, and $\bm{\lambda}^{\left(v\right)}$ into \eqref{bq2}, the constraints in \eqref{bq2} are
\begin{align}
\label{cq4}
2 \mbox{Re}\left\{\bm{\mu}_{k,1}^H\bm{\phi}^* \right\}+\bm{\phi}^T \bm{\Upsilon}_{k,1} \bm{\phi}^*\leq e_{k,1},\ k\in\mathcal{K}
\end{align}
where $e_{k,1}= t_{k,1}^{(v)} -\tau_{k,c} - | u_{k,1}^{(v)}|^2  (\sigma^2+\psi_{k,c}^{(v)})$,
\begin{align}
\bm{\mu}_{k,1} &= |u_{k,1}^{\left(v\right)}|^2\mathbf{G}_{k,c}\mathbf{H}\bm{\Sigma}_k\mathbf{h}_{k,c} -(u_{k,1}^{\left(v\right)})^H \mathbf{G}_{k,c}\mathbf{H}\mathbf{w}_{k,c}^{\left(v\right)}, \\
\bm{\Sigma}_k &= \sum_{i \neq k}^{K}\left(\mathbf{W}_{i,c}^{\left(v\right)}+ \mathbf{W}_{i,e}^{\left(v\right)}\right), \mathbf{W}_{k,i}^{\left(v\right)}=\mathbf{w}_{k,i}^{\left(v\right)}(\mathbf{w}_{k,i}^{\left(v\right)})^H,\\
\bm{\Upsilon}_{k,1} &= |u_{k,1}^{\left(v\right)}|^2\mathbf{G}_{k,c}\mathbf{H} \bm{\Sigma}_k\mathbf{H}^H\mathbf{G}_{k,c}^H, \\
t_{k,1}^{\left(v\right)}&=2\mbox{Re}\left\{(u_{k,1}^{\left(v\right)})^H\mathbf{h}_{k,c}^H\mathbf{w}_{k,c}^{\left(v\right)} \right\},\\
\psi_{k,i}^{\left(v\right)}&=\sum_{j=1, j\neq k}^{K}\left(|\mathbf{h}_{k,i}^H\mathbf{w}_{j,c}^{\left(v\right)}|^2 + |\mathbf{h}_{k,i}^H\mathbf{w}_{j,e}^{\left(v\right)}|^2\right).
\end{align}
Similarly, the constraints in \eqref{bq3} and \eqref{bq4} are
\begin{align}
\label{cq10}
2 \mbox{Re}\left\{\bm{\mu}_{k,2}^H\bm{\phi}^* \right\}+\bm{\phi}^T \bm{\Upsilon}_{k,2} \bm{\phi}^*\leq e_{k,2},\ k\in\mathcal{K},\\
\label{cq11}2 \mbox{Re}\left\{\bm{\mu}_{k,3}^H\bm{\phi}^* \right\}+\bm{\phi}^T \bm{\Upsilon}_{k,3} \bm{\phi}^*\leq e_{k,3},\ k\in\mathcal{K},
\end{align}
where $e_{k,2}= t_{k,2}^{(v)} -\tau_{k,e} - |u_{k,2}^{(v)}|^2  (\sigma^2+\hat{\psi}_{k,e}^{(v)})$, $e_{k,3}= t_{k,3}^{(v)} -\tau_{k,e} - | u_{k,3}^{(v)}|^2  (\sigma^2+\hat{\psi}_{k,c}^{(v)})$,
\begin{align}
\bm{\mu}_{k,2} &=|u_{k,2}^{\left(v\right)}|^2 \mathbf{G}_{k,e}\mathbf{H}\hat{\bm{\Sigma}}_k\mathbf{h}_{k,e} - (u_{k,2}^{\left(v\right)})^H\mathbf{G}_{k,e}\mathbf{H}\mathbf{w}_{k,e}^{\left(v\right)},\\
\bm{\mu}_{k,3} &=|u_{k,3}^{\left(v\right)}|^2 \mathbf{G}_{k,c}\mathbf{H}\hat{\bm{\Sigma}}_k\mathbf{h}_{k,c} - (u_{k,3}^{\left(v\right)})^H\mathbf{G}_{k,c}\mathbf{H}\mathbf{w}_{k,e}^{\left(v\right)},\\
\bm{\Upsilon}_{k,2} &=|u_{k,2}^{\left(v\right)}|^2\mathbf{G}_{k,e}\mathbf{H} \hat{\bm{\Sigma}}_k\mathbf{H}^H\mathbf{G}_{k,e}^H, \\
\bm{\Upsilon}_{k,3} &=|u_{k,3}^{\left(v\right)}|^2\mathbf{G}_{k,c}\mathbf{H} \hat{\bm{\Sigma}}_k\mathbf{H}^H\mathbf{G}_{k,c}^H, \\
\hat{\bm{\Sigma}}_k &=\mathbf{W}_{k,c}^{\left(v\right)}+\bm{\Sigma}_k,\ \hat{\psi}_{k,i}^{\left(v\right)}=\psi_{k,i}^{\left(v\right)}+|\mathbf{h}_{k,i}^H\mathbf{w}_{k,c}^{\left(v\right)}|^2,\\
t_{k,2}^{\left(v\right)}&=2\mbox{Re}\left\{(u_{k,2}^{\left(v\right)})^H\mathbf{h}_{k,e}^H\mathbf{w}_{k,e}^{\left(v\right)} \right\},\\
t_{k,3}^{\left(v\right)}&=2\mbox{Re}\left\{(u_{k,3}^{\left(v\right)})^H\mathbf{h}_{k,c}^H\mathbf{w}_{k,e}^{\left(v\right)} \right\}.
\end{align}
From \eqref{cq4}, \eqref{cq10}, and \eqref{cq11}, problem \eqref{cq2} is a QCQP problem which can be equivalently transformed into the following SOCP problem
\begin{align}\label{cq20}
\min_{\bm{\phi}} \ & \left\|\bm{\phi} - \bm{\varphi}^{\left(v\right)}  + \bm{\lambda}^{\left(v\right)} \right\|_2 \nonumber\\
\mbox{s.t.}\ \ & \left\|\!\!\begin{array}{c}
\bm{\Upsilon}_{k,j}^{\frac{1}{2}} \bm{\phi}^* \\
\frac{1-e_{k,j}}{2}+\mbox{Re}\left\{\bm{\mu}_{k,j}^H\bm{\phi}^*\right\}\end{array}\!\!\right\|_2\!\! \leq \frac{1+e_{k,j}}{2}-\mbox{Re}\left\{\bm{\mu}_{k,j}^H\bm{\phi}^*\right\}, \nonumber \\ & \quad \quad k \in \mathcal{K},\ j\in \{1,2,3\}.
\end{align}
Since the SOCP problem \eqref{cq20} is a convex problem, it can be efficiently solved by using some off-the-shelf convex optimization tools, e.g., CVX \cite{MGrant}.

\textit{2) Update $\mathbf{w}$}:
The sub-problem for updating variable $\mathbf{w}$ is expressed as
\begin{align} \label{cq21}
\mathbf{w}^{\left(v+1\right)} = \arg	\underset{\mathbf{w}}{\min} \mathcal{L}_{\xi}\left(\bm{\phi}^{\left(v+1\right)}, \mathbf{w}, \mathbf{u}^{\left(v\right)}, \bm{\varphi}^{\left(v\right)}, \bm{\lambda}^{\left(v\right)}\right).
\end{align}
The above update of $\mathbf{w}$ in problem \eqref{cq21} is equivalent to solving following QCQP problem
\begin{align}
\label{cq22}
\min_{\mathbf{w}} \ & \left\|\mathbf{w} \right\|_2^2 \nonumber\\
\mbox{s.t.}\ & \mathbf{w}^H \bm{\Psi}_{k,j} \mathbf{w} + 2 \mbox{Re}\left\{\bm{\rho}_{k,j}^H\mathbf{w} \right\}  \leq \hat{e}_{k,j}, \nonumber \\ & \quad \quad k \in \mathcal{K},\ j\in \{1,2,3\}
\end{align}
where $\hat{e}_{k,1}=-\tau_{k,c}-|u_{k,1}^{\left(v\right)}|^2\sigma^2$, $\hat{e}_{k,2}=-\tau_{k,e}-|u_{k,2}^{\left(v\right)}|^2\sigma^2$, $\hat{e}_{k,3}=-\tau_{k,e}-|u_{k,3}^{\left(v\right)}|^2\sigma^2$,
\begin{align}
\bm{\Psi}_{k,1}&=|u_{k,1}^{\left(v\right)}|^2 \mbox{Bdiag}
\begin{bmatrix}
\smash{\underbrace{\mathbf{V}_{k,c}, \cdots, \mathbf{V}_{k,c}}_{2k-2}, \mathbf{0}, \mathbf{0},\underbrace{\mathbf{V}_{k,c}, \cdots, \mathbf{V}_{k,c}}_{2K-2k}}
\end{bmatrix},\\
\bm{\Psi}_{k,2} &=|u_{k,2}^{\left(v\right)}|^2\mbox{Bdiag}
\begin{bmatrix}
\smash{\underbrace{\mathbf{V}_{k,e}, \cdots, \mathbf{V}_{k,e}}_{2k-1}, \mathbf{0}, \underbrace{\mathbf{V}_{k,e}, \cdots, \mathbf{V}_{k,e}}_{2K-2k}}
\end{bmatrix}, \\
\bm{\Psi}_{k,3} &=|u_{k,3}^{\left(v\right)}|^2\mbox{Bdiag}
\begin{bmatrix}
\smash{\underbrace{\mathbf{V}_{k,c}, \cdots, \mathbf{V}_{k,c}}_{2k-1}, \mathbf{0},\underbrace{\mathbf{V}_{k,c}, \cdots, \mathbf{V}_{k,c}}_{2K-2k}}
\end{bmatrix},\\
\mathbf{V}_{k,i}&= \hat{\mathbf{h}}_{k,i}^{\left(v+1\right)}(\hat{\mathbf{h}}_{k,i}^{\left(v+1\right)})^H,\\ \bm{\rho}_{k,1}&= u_{k,1}^{\left(v\right)}\hat{\mathbf{h}}_{k,c}^{\left(v+1\right)},\ \hat{\mathbf{h}}_{k,i}^{\left(v+1\right)}=\mathbf{H}^H\mathbf{G}_{k,i}^H{\bm{\phi}^{\left(v+1\right)} }^* + \mathbf{h}_{k,i}^{\left(v+1\right)},\\
\bm{\rho}_{k,2}&= u_{k,2}^{\left(v\right)}\hat{\mathbf{h}}_{k,e}^{\left(v+1\right)},\
\bm{\rho}_{k,3}= u_{k,3}^{\left(v\right)}\hat{\mathbf{h}}_{k,c}^{\left(v+1\right)}.
\end{align}
Therefore, the QCQP problem \eqref{cq22} can be rewritten as the following SOCP problem
\begin{align}\label{cq31}
\min_{\mathbf{w}} \ & \left\|\mathbf{w} \right\|_2 \nonumber\\
\mbox{s.t.}\ & \left\|\!\!\begin{array}{c}
\bm{\Psi}_{k,j}^{\frac{1}{2}} \mathbf{w} \\
\frac{1-\hat{e}_{k,j}}{2}+\mbox{Re}\left\{\bm{\rho}_{k,j}^H\mathbf{w} \right\}
\end{array}\!\!\right\|_2\!\! \leq \frac{1+\hat{e}_{k,j}}{2}-\mbox{Re}\left\{\bm{\rho}_{k,j}^H\mathbf{w} \right\}, \nonumber \\ & \quad \quad k \in \mathcal{K},\ j \in \{1,2,3\}.
\end{align}

\textit{3) Update $\mathbf{u}$}: Based on Lemma 1, the update of each element in $\mathbf{u}$ can be given as
\begin{align} \label{cq32}
u_{k,1}^{\left(v+1\right)} & = \frac{(\hat{\mathbf{h}}_{k,c}^{\left(v+1\right)})^H \mathbf{w}_{k,c}^{\left(v+1\right)}}{\sigma^2 + \zeta_{k,c}^{\left(v+1\right)}},  \\
\label{cq33}
u_{k,2}^{\left(v+1\right)} & = \frac{(\hat{\mathbf{h}}_{k,e}^{\left(v+1\right)})^H \mathbf{w}_{k,e}^{\left(v+1\right)}}{ \sigma^2 + |(\hat{\mathbf{h}}_{k,e}^{\left(v+1\right)})^H \mathbf{w}_{k,c}^{\left(v+1\right)}|^2 + \zeta_{k,e}^{\left(v+1\right)}}, \\
\label{cq34}
u_{k,3}^{\left(v+1\right)} & = \frac{(\hat{\mathbf{h}}_{k,c}^{\left(v+1\right)})^H \mathbf{w}_{k,e}^{\left(v+1\right)}}{\sigma^2 + |(\hat{\mathbf{h}}_{k,c}^{\left(v+1\right)})^H \mathbf{w}_{k,c}^{\left(v+1\right)}|^2 + \zeta_{k,c}^{\left(v+1\right)}},
\end{align}
where
\begin{equation}
\zeta_{k,i}^{\left(v+1\right)}=\sum_{j=1, j\neq k}^{K}\left(|(\hat{\mathbf{h}}_{k,i}^{\left(v+1\right)})^H\mathbf{w}_{j,c}|^2 + |(\hat{\mathbf{h}}_{k,i}^{\left(v+1\right)})^H\mathbf{w}_{j,e}|^2\right).
\end{equation}

\textit{4) Update $\bm{\varphi}$}: the update of  $\bm{\varphi}$ can be expressed as
\begin{align} \label{cq36}
\bm{\varphi}^{\left(v+1\right)} = \arg \min_{\bm{\varphi} \in \mathcal{F}} \left\|\bm{\phi}^{\left(v+1\right)} + \bm{\lambda}^{\left(v\right)}  - \bm{\varphi}\right\|_2^2.
\end{align}
The goal of problem \eqref{cq36} is to project $\bm{\phi}^{\left(v+1\right)} + \bm{\lambda}^{\left(v\right)}$ onto the set $\mathcal{F}$. Let $\phi_m^{\left(v+1\right)}$ and $\lambda_m^{\left(v\right)}$ denote the $m$-th elements of $\bm{\phi}^{\left(v+1\right)}$ and $\bm{\lambda}^{\left(v\right)}$, respectively. Based on the three different assumptions for $\bm{\phi}$, $\bm{\varphi}$ can be updated as follows:

If $\mathcal{F} = \left\{\bm{\phi} \big| |\phi_m|^2 \in \left[0,1\right] \right\}$, the $m$-th element of $\bm{\varphi}^{\left(v+1\right)}$ can be obtained by solving following problem
\begin{align}
\min_{\varphi_m} & \quad \left|\phi_m^{\left(v+1\right)} + \lambda_m^{\left(v\right)}-\varphi_m \right| \nonumber \\
	\text{s.t.} & \quad \ \varphi_m^H \varphi_m \leq 1,  \quad m \in \{1,\cdots, M\}.
\end{align}
Hence, we have
\begin{align}\label{cq38}
\varphi_m^{\left(v+1\right)} =\begin{cases}
\phi_m^{\left(v+1\right)}+\lambda_m^{\left(v\right)}; &  \mbox{if} \ \left| \phi_m^{\left(v+1\right)}+\lambda_m^{\left(v\right)} \right| \leq 1, \\
\frac{\phi_m^{\left(v+1\right)}+\lambda_m^{\left(v\right)}}{\left|\phi_m^{\left(v+1\right)}+\lambda_m^{\left(v\right)}\right|}; & \mbox{otherwise}.
\end{cases}
\end{align}

If $\mathcal{F} = \left\{\bm{\phi} \big| |\phi_m|^2 = 1 \right\}$,
\begin{align} \label{cq39}
\varphi_m^{\left(v+1\right)}= \frac{\phi_m^{\left(v+1\right)}+\lambda_m^{\left(v\right)}}{\left|\phi_m^{\left(v+1\right)}+\lambda_m^{\left(v\right)}\right|}.
\end{align}

If $\mathcal{F} = \left\{\bm{\phi} \big| |\phi_m|^2 = 1, \theta_m\in\mathcal{A}\right\}$,
\begin{align} \label{cq40}
\varphi_m^{\left(v+1\right)}= e^{j \frac{2\pi}{L}\scriptsize\mbox{Round}\left(\frac{\angle \left(\phi_m^{\left(v+1\right)}+\lambda_m^{\left(v\right)}\right)}{\frac{2\pi}{L}}\right)}.
\end{align}

\textit{5) Update $\bm{\lambda}$}: the update of the dual variable $\bm{\lambda}$ can be given as follows
\begin{align} \label{cq41}
\bm{\lambda}^{\left(v+1\right)} = \bm{\lambda}^{\left(v\right)} + \bm{\phi}^{\left(v+1\right)} - \bm{\varphi}^{\left(v+1\right)}.
\end{align}
The whole algorithm of our proposed SOCP-ADMM based beamforming design is summarized in Algorithm 1.
\begin{algorithm}[H]
\caption{Proposed SOCP-ADMM Based Algorithm.}
\begin{algorithmic}[1]
\STATE\textbf{Initialize:} $v= 0$, $\bm{\phi}^{(0)}$, $\mathbf{w}^{\left(0\right)}$, $\mathbf{u}^{\left(0\right)}$, $\bm{\varphi}^{\left(0\right)}$, $\bm{\lambda}^{\left(0\right)}$;
\STATE \textbf{Repeat:} \\
Update $\bm{\phi}^{\left(v+1\right)}$ by solving SOCP problem \eqref{cq20};\\
Update $\mathbf{w}^{\left(v+1\right)}$ by solving SOCP problem \eqref{cq31};\\
Update $\mathbf{u}^{\left(v+1\right)}$ by using closed-form solutions \eqref{cq32}, \eqref{cq33} and \eqref{cq34}; \\
Depending on different assumptions for reflection coefficients, update $\bm{\varphi}^{\left(v+1\right)}$ according to \eqref{cq38}, \eqref{cq39}, and \eqref{cq40}, respectively; \\
Update Lagrange multiplier $\bm{\lambda}^{\left(v+1\right)}$ by \eqref{cq41};\\
$v:=v+1$;
\STATE \textbf{Until:} Convergence.
\end{algorithmic}
\end{algorithm}
\textit{Remark 1:} The computational complexity of solving an SOCP problem is \cite{MBeko,IPolik}
\begin{align}
\mathcal{O}\left(k_{soc}^{0.5}\left(m_{soc}^3 + m_{soc}^2 \sum_{i = 1}^{k_{soc}}n_{i,soc} + \sum_{i = 1}^{k_{soc}}n_{i,soc}^2\right)\right),
\end{align}
where $k_{soc}$ denotes the number of second-order cone (SOC) constraints, $m_{soc}$ denotes the dimension of the optimization variable, and $n_{i,soc}$ denotes the dimension of the $i$-th SOC. For the first SOCP sub-problem \eqref{cq20} in our paper, we have $k_{soc} = 3K + 1$, $m_{soc} = M$, $n_i = M+1$. Therefore, the complexity of solving sub-problem \eqref{cq20} is $\mathcal{O}\left(K^{1.5}M^3\right)$. Similarly, for the second SOCP sub-problem \eqref{cq31}, we have $k_{soc} = 3K + 1$, $m_{soc} = 2KN$, $n_i = 2NK+1$. Therefore, the complexity of solving sub-problem \eqref{cq31} is $\mathcal{O}\left(K^{4.5}N^3\right)$. Based on the rule of matrix multiplication operation, the complexity of updating $\mathbf{u}$ in \eqref{cq32}$-$\eqref{cq34} is $\mathcal{O}\left(N \left(2K-1\right)K + 2K^2N + 2K^2N\right) = \mathcal{O}\left(NK^2\right)$. The complexities of updating $\bm{\varphi}$ and $\bm{\lambda}$ are both $\mathcal{O}\left(M\right)$. Hence, the overall complexity of our proposed SOCP-ADMM method above within an accuracy $\epsilon$ is $\mathcal{O}\left(\left(K^{1.5}M^3 + K^{4.5}N^3\right)\cdot \log \left(1/\epsilon\right)\right)$.

\section{ZF Based Suboptimal Algorithm}

To reduce the computational complexity, in this section, we propose a low complexity ZF based suboptimal algorithm. Based on the ZF scheme \cite{QHSpencer},
the beamforming vectors $\mathbf{w}_{k,e}$ and $\mathbf{w}_{k,c}$ are designed to eliminate inter-cluster interferences. Therefore, we impose the constraint
\begin{align}\label{dq1}
\hat{\mathbf{h}}_{k_1,i_1}\mathbf{w}_{k_2,i_2} = 0
\end{align}
for $k_1\neq k_2$, $k_1\in \mathcal{K}$, $k_2\in \mathcal{K}$, $i_1\in \{c,e\}$, $i_2\in \{c,e\}$. To make \eqref{dq1} possible, we assume that $N \geq  2K - 1$. Define
\begin{align}
\bar{\mathbf{H}}_k = \left[\hat{\mathbf{H}}_1, \cdots, \hat{\mathbf{H}}_{k-1},
\hat{\mathbf{H}}_{k+1}, \cdots , \hat{\mathbf{H}}_K \right]
\end{align}
where $\hat{\mathbf{H}}_k=[\hat{\mathbf{h}}_{k,e}, \hat{\mathbf{h}}_{k,c}]$. The singular value decomposition (SVD) of $\bar{\mathbf{H}}_k$ can be formulated as
\begin{align}\label{dq3}
\bar{\mathbf{H}}_k = \bar{\mathbf{U}}_k\bar{\bm{\Sigma}}_k\bar{\mathbf{V}}_k^H
\end{align}
where $\bar{\mathbf{U}}_k \in \mathbb{C}^{N \times N}$ and $\bar{\mathbf{V}}_{k} \in \mathbb{C}^{\left(2K-2\right) \times  \left(2K - 2\right)}$ are unitary matrices, $\bar{\bm{\Sigma}}_{k} \in \mathbb{C}^{N \times 2\left(K-1\right)}$ is a diagonal matrix. After applying the ZF beamforming scheme, we have
\begin{align}
\mathbf{w}_{k,i} = \mathbf{U}_k \bar{\mathbf{w}}_{k,i}, \ i \in \{e,c\}
\end{align}
where $\mathbf{U}_k$ consists of the last $N - 2K + 2$ columns of $\bar{\mathbf{U}}_k$ and $\bar{\mathbf{w}}_{k,i}$ is the vector to be optimized. Therefore, $\mathbf{U}_k$ lies in the null space of $\bar{\mathbf{H}}_k$, which results in ${\mathbf{U}}_k^H\bar{\mathbf{H}}_k = \mathbf{0}$ and that \eqref{dq1} is satisfied.
	
Using \eqref{dq1}, the expressions $\gamma_{k,e}$, $\gamma_{k,c,e}$ and $\gamma_{k,c}$ in \eqref{q5}, \eqref{q7} and \eqref{q8} can be reformulated as
\begin{align}
\hat{\gamma}_{k,e}&=  \frac{|\hat{\mathbf{h}}_{k,e}^H \mathbf{w}_{k,e}|^2}{\sigma^2 + |\hat{\mathbf{h}}_{k,e}^H\mathbf{w}_{k,c}|^2}, \\
\hat{\gamma}_{k,c,e} &= \frac{|\hat{\mathbf{h}}_{k,c}^H \mathbf{w}_{k,e}|^2}{\sigma^2 + |\hat{\mathbf{h}}_{k,c}^H\mathbf{w}_{k,c}|^2}, \\
\hat{\gamma}_{k,c} &=|\hat{\mathbf{h}}_{k,c}^H \mathbf{w}_{k,c}|^2 /\sigma^2.
\end{align}
Thus, the transmit power minimization problem in \eqref{q9} can be recast as
\begin{subequations}\label{dq9}
\begin{align}
\label{dq9a}\min_{\bm{\phi}, \{\bar{\mathbf{w}}_{k,i}\}}\ & \sum_{k = 1}^{K} \left( \left\|\mathbf{U}_k\bar{\mathbf{w}}_{k,c}\right\|^2+ \left\|\mathbf{U}_k\bar{\mathbf{w}}_{k,e}\right\|^2\right) \\
\label{dq9b} \mbox{s.t.}\ \ & \hat{\gamma}_{k,c} \geq \tau_{k,c}, \ \forall\ k\in\mathcal{K},\\
\label{dq9c} & \min\left(\hat{\gamma}_{k,e}, \hat{\gamma}_{k,c,e}\right) \geq \tau_{k,e}, \ \forall\ k\in\mathcal{K},\\
\label{dq9d} &  \bm{\phi} \in \mathcal{F}.
\end{align}
\end{subequations}
Problem \eqref{dq9} is non-convex and alternating optimization is proposed to iteratively optimize $\{\bar{\mathbf{w}}_{k,i}\}$ and $\bm{\phi}$.
	
\subsection{Optimization of $\{\bar{\mathbf{w}}_{k,c}$, $\bar{\mathbf{w}}_{k,e}\}$}
Given $\bm{\phi}$, from \eqref{dq9}, the optimization of $\left\{\bar{\mathbf{w}}_{k,c},\bar{\mathbf{w}}_{k,e}\right\}$, $k \in \mathcal{K}$ can be decoupled into $K$ optimization problems. In the $k$-th cluster, the transmit power minimization problem is reformulated as
\begin{subequations}\label{dq10}
\begin{align}
\label{dq10a}\min_{\bar{\mathbf{w}}_{k,c},\bar{\mathbf{w}}_{k,e}}\ &\left\|\bar{\mathbf{w}}_{k,c}\right\|^2+ \left\|\bar{\mathbf{w}}_{k,e}\right\|^2 \\
\label{dq10b} \mbox{s.t.}\ \ & \left|\bar{\mathbf{h}}_{k,c}^H \bar{\mathbf{w}}_{k,c} \right|^2 \geq \tau_{k,c} \sigma^2,\\
\label{dq10c} & \left|\bar{\mathbf{h}}_{k,e}^H \bar{\mathbf{w}}_{k,e}\right|^2 \geq \tau_{k,e}\left(\sigma^2 + \left|\bar{\mathbf{h}}_{k,e}^H\bar{\mathbf{w}}_{k,c}\right|^2\right), \\
\label{dq10d} &  \left|\bar{\mathbf{h}}_{k,c}^H\bar{\mathbf{w}}_{k,e} \right|^2 \geq \tau_{k,e}\left(\sigma^2 + \left|\bar{\mathbf{h}}_{k,c}^H\bar{\mathbf{w}}_{k,c} \right|^2\right),
\end{align}
\end{subequations}
where $\bar{\mathbf{h}}_{k,c} = \hat{\mathbf{h}}_{k,c}\mathbf{U}_k$ and $\bar{\mathbf{h}}_{k,e} = \hat{\mathbf{h}}_{k,e}\mathbf{U}_k$.

We propose to optimize $\bar{\mathbf{w}}_{k,c}$ by neglecting constraints \eqref{dq10c} and \eqref{dq10d} to obtain a suboptimal closed-form solution of $\bar{\mathbf{w}}_{k,c}$. By neglecting constraints \eqref{dq10c} and \eqref{dq10d}, the optimization problem of $\bar{\mathbf{w}}_{k,c}$ is reduced to
\begin{align}\label{dq11}
\min_{\bar{\mathbf{w}}_{k,c}}\ &\left\|\bar{\mathbf{w}}_{k,c}\right\|^2 \nonumber\\
\mbox{s.t.}\ \ & \left|\bar{\mathbf{h}}_{k,c}^H \bar{\mathbf{w}}_{k,c} \right|^2 \geq \tau_{k,c} \sigma^2
\end{align}
whose solution is
\begin{align} \label{dq12}
\bar{\mathbf{w}}_{k,c}^\star =\tau_{k_c}^{\frac{1}{2}} \sigma \frac{\bar{\mathbf{h}}_{k,c}}{\left\|\bar{\mathbf{h}}_{k,c}\right\|^2}.
\end{align}

When $\bar{\mathbf{w}}_{k,c}^\star$ is obtained, the optimization problem of $\bar{\mathbf{w}}_{k,e}$ is reduced to
\begin{subequations}\label{dq13}
\begin{align}
\label{dq13a}
\min_{\bar{\mathbf{w}}_{k,e}}\ &\left\|\bar{\mathbf{w}}_{k,e}\right\|^2 \\
\label{dq13b} \mbox{s.t.}\ \ & \left|\mathbf{b}_e^H \bar{\mathbf{w}}_{k,e}\right|^2 \geq 1, \\
\label{dq13c} &  \left|\mathbf{b}_c^H\bar{\mathbf{w}}_{k,e} \right|^2 \geq 1,
\end{align}
\end{subequations}
where
\begin{align}
\label{dq14}
\mathbf{b}_e&=\bar{\mathbf{h}}_{k,e}\tau_{k,e}^{-\frac{1}{2}}\left(\sigma^2 + \left|\bar{\mathbf{h}}_{k,e}^H\bar{\mathbf{w}}_{k,c}^\star \right|^2 \right)^{-\frac{1}{2}}, \\
\label{dq15}
\mathbf{b}_c&=\bar{\mathbf{h}}_{k,c}\tau_{k,e}^{-\frac{1}{2}}\left(\sigma^2 + \left|\bar{\mathbf{h}}_{k,c}^H\bar{\mathbf{w}}_{k,c}^\star \right|^2 \right)^{-\frac{1}{2}}.
\end{align}

If $\mathbf{b}_e\mathbf{b}_e^H - \mathbf{b}_c\mathbf{b}_c^H \succeq \mathbf{0}$, the constraint \eqref{dq13c} is active and the optimal closed-form solution of $\bar{\mathbf{w}}_{k,e}$ is $\bar{\mathbf{w}}_{k,e}^\star = \frac{\mathbf{b}_c}{\left\| \mathbf{b}_c\right\|_2^2}$.
	
If $\mathbf{b}_e\mathbf{b}_e^H - \mathbf{b}_c\mathbf{b}_c^H  \preceq \mathbf{0}$, the constraint \eqref{dq13b} is active and the optimal closed-form solution of $\bar{\mathbf{w}}_{k,e}$ is $\bar{\mathbf{w}}_{k,e}^\star = \frac{\mathbf{b}_e}{\left\| \mathbf{b}_e\right\|_2^2}$.
	
If $\mathbf{b}_e\mathbf{b}_e^H - \mathbf{b}_c\mathbf{b}_c^H$ is indefinite, the optimal closed-form solution of $\bar{\mathbf{w}}_{k,e}$ is \cite{JChoi}
\begin{equation}
\bar{\mathbf{w}}_{k,e}^\star = \frac{\mathbf{b}_e + f\left(\theta^\star\right)\mathbf{b}_c}{\left\|\mathbf{b}_e + f\left(\theta^\star\right)\mathbf{b}_c\right\|_2^2}
\end{equation}
where
\begin{equation}
f\left(x\right) = \frac{\left\|\mathbf{b}_e\right\|^2 - e^{j x} \mathbf{b}_c^H\mathbf{b}_e}{e^{j x}\left\|\mathbf{b}_c\right\|^2 - \mathbf{b}_e^H\mathbf{b}_c}
\end{equation}
and
\begin{align}\label{dq18}
\theta^\star = \arg \max_{\theta}\left| \left\| \mathbf{b}_e\right\|^2 + \mathbf{b}_e^H\mathbf{b}_cf\left(\theta\right) \right|^2.
\end{align}
The optimal $\theta^\star$ in \eqref{dq18} can be obtained by performing a one-dimensional search over $[0, 2 \pi)$.

\subsection{Optimization of $\bm{\phi}$}
Given $\bar{\mathbf{w}}_{k,e}$ and $\bar{\mathbf{w}}_{k,c}$, $k \in \mathcal{K}$, through observations, we propose an efficient suboptimal solution of $\bm{\phi}$.

In problem \eqref{dq9}, since the objective function does not includes $\bm{\phi}$, we propose to optimize $\bm{\phi}$ such that the sum received signal power at the central and cell-edge users, are maximized, i.e.,
\begin{align}\label{dq19}
\max_{\bm{\phi}\in \mathcal{F}}\ &\sum_{k= 1}^{K} \left(\left|\bar{\mathbf{h}}_{k,c}^H \bar{\mathbf{w}}_{k,c} \right|^2+ \left|\bar{\mathbf{h}}_{k,e}^H \bar{\mathbf{w}}_{k,e}\right|^2+\left|\bar{\mathbf{h}}_{k,c}^H\bar{\mathbf{w}}_{k,e} \right|^2\right).
\end{align}
After some mathematical manipulations, we rewrite problem \eqref{dq19} as follows
\begin{align}\label{dq20}
\max_{\bm{\phi}\in \mathcal{F}}\ & \sum_{k= 1}^{K}\sum_{j=1}^{3} \left|\bm{\phi}^T\bm{\varpi}_{k,j} +\varsigma_{k,j}\right|^2.
\end{align}
where
\begin{align}
\bm{\varpi}_{k,1}&= \mathbf{G}_{k,c}\mathbf{H}\mathbf{w}_{k,c},\ \varsigma_{k,1}=\mathbf{h}_{k,c}^H\mathbf{w}_{k,c}\\
\bm{\varpi}_{k,2}&= \mathbf{G}_{k,e}\mathbf{H}\mathbf{w}_{k,e},\ \varsigma_{k,2}=\mathbf{h}_{k,e}^H\mathbf{w}_{k,e}\\ \bm{\varpi}_{k,3} &= \mathbf{G}_{k,c}\mathbf{H}\mathbf{w}_{k,e},\ \varsigma_{k,3}=\mathbf{h}_{k,c}^H\mathbf{w}_{k,e}
\end{align}
Problem \eqref{dq20} can be equivalently rewritten as
\begin{align}\label{dq24}
\max_{\tilde{\bm{\phi}}\in \mathcal{F}}\ & \tilde{\bm{\phi}}^H\bm{\Omega}\tilde{\bm{\phi}}
\end{align}
where $\tilde{\bm{\phi}}=[\bm{\phi}^H, 1]^T$, $\bm{\Omega}= \sum_{k = 1}^{K}\sum_{j = 1}^{3}\bm{\Omega}_{k,j}$, and
\begin{align}
\mathbf{\Omega}_{k,j} &= \begin{bmatrix}
\bm{\varpi}_{k,j}\bm{\varpi}_{k,j}^H & \bm{\varpi}_{k,j}\varsigma_{k,j}^* \\
\bm{\varpi}_{k,j}^H\varsigma_{k,j} & 0
\end{bmatrix}.
\end{align}

If $\mathcal{F} = \left\{\bm{\phi} \big| |\phi_m|^2 \in \left[0,1\right] \right\}$ or $\mathcal{F} = \left\{\bm{\phi} \big| |\phi_m|^2 = 1 \right\}$,
using the fixed point iteration method proposed in \cite{XYu}, a locally optimal solution to problem \eqref{dq24} is obtained in an iterative manner. In the $v$-th iteration, we update
\begin{align}\label{dq27}
\tilde{\bm{\phi}}^{\left(v+1\right)} = \left(\bm{\Omega}\tilde{\bm{\phi}}^{\left(v\right)}\right)\oslash \mbox{Abs}\left(\bm{\Omega}\tilde{\bm{\phi}}^{\left(v\right)}\right)
\end{align}
The convergence of the aforementioned fixed point iterations is proven in \cite{XYu}. After convergence, we obtain a locally optimal solution to problem \eqref{dq24} as follows
\begin{equation}\label{dq28}
\tilde{\bm{\phi}}^\star=\frac{\tilde{\bm{\phi}}^{\left(v\right)}}{\tilde{\phi}_{M+1}^{\left(v\right)}}
\end{equation}
where $\tilde{\phi}_{M+1}^{\left(v\right)}$ denotes the entry in the $(M+1)$-th  column of $\tilde{\bm{\phi}}^{\left(v\right)}$, which ensures that the entry in the $(M+1)$-th column of $\tilde{\bm{\phi}}^\star$ is 1.

If $\mathcal{F} = \left\{\bm{\phi} \big| |\phi_m|^2 = 1, \theta_m\in\mathcal{A}\right\}$, the locally optimal solution to problem \eqref{bq12} is
\begin{align}\label{dq29}
e^{j\frac{2\pi}{L}\mbox{Round}\left(\frac{L\angle \tilde{\bm{\phi}}^\star}{2\pi}\right)}.
\end{align}

The whole algorithm of our proposed ZF based suboptimal beamforming design is summarized in Algorithm 2.
\begin{algorithm}[H]
\caption{Proposed ZF Based Suboptimal Algorithm.}
\begin{algorithmic}[1]
\STATE \textbf{Repeat:} \\
Update $\bar{\mathbf{w}}_{k,c}$ and $\bar{\mathbf{w}}_{k,e}$, $\forall\ k \in \mathcal{K}$;
\STATE \textbf{Repeat:} \\
Update $\tilde{\bm{\phi}}$ using \eqref{dq27}; \\
\STATE \textbf{Until:} Convergence; \\
Obtain $\tilde{\bm{\phi}}^\star$ using \eqref{dq28};\\
If $\mathcal{F} = \left\{\bm{\phi} \big| |\phi_m|^2 = 1, \theta_m\in\mathcal{A}\right\}$, modify the locally optimal solution using \eqref{dq29};\\
\STATE \textbf{Until:} Convergence.
\end{algorithmic}
\end{algorithm}

\textit{Remark 2:} From Algorithm 2, the computational complexity of obtaining $\bar{\mathbf{w}}_{k,c}$ and $\bar{\mathbf{w}}_{k,e}$ mainly comes from the SVD in \eqref{dq3}. For a matrix $\mathbf{A} \in \mathbb{C}^{a\times b}$, the complexity of calculating SVD is $\mathcal{O}(ab^2)$ \cite{GHGolub}. Thus, the complexity here is $\mathcal{O}(N(2K-2)^2)$. The complexity of obtaining $\bm{\phi}$ involves matrix multiplication in \eqref{dq27}, which is $\mathcal{O}\left(L_1M^2\right)$, where $L_1$ denotes the number of fixed point iterations. Therefore, the total complexity of ZF based suboptimal algorithm is $\mathcal{O}\left(L_2N(2K-2)^2+L_2L_1M^2\right)$, where $L_2$ denotes the number of  alternating optimization iterations of $(\bar{\mathbf{w}}_{k,c}, \bar{\mathbf{w}}_{k,e})$ and $\bm{\phi}$.

\section{Simulation Results}

In simulations, we assume that the number of clusters is $K = 3$. The channels are generated as follows
\begin{align}
\mathbf{H}&= C_0d_0^{-\alpha_0}\breve{\mathbf{H}},\\
\mathbf{g}_{k,i}&= C_0d_{k,i}^{-\alpha_{k,i}}\breve{\mathbf{g}}_{k,i},\\
\mathbf{h}_{k,i}&= C_0\tilde{d}_{k,i}^{-\tilde{\alpha}_{k,i}}\breve{\mathbf{h}}_{k,i},
\end{align}
for $k \in \mathcal{K}$, $i\in \{c,e\}$, where $C_0 = -30$dB denotes the path loss at the reference distance of one meter; $d_0=30$ m, $d_{k,c}=50$ m, $d_{k,e} = 70$ m, $\tilde{d}_{k,c}=50$ m, and $\tilde{d}_{k,e} = 80$ m denote the distances from the BS to the IRS, from the IRS to $U_{k,c}$ and $U_{k,e}$, and from the BS to $U_{k,c}$ and $U_{k,e}$, respectively; $\alpha_{0}=\alpha_{k,i}=2.5$ and $\tilde{\alpha}_{k,i} = 3.5$ denote the corresponding path loss exponents; all the entries in $\breve{\mathbf{g}}_{k,i}$, $\breve{\mathbf{h}}_{k,i}$ and $\breve{\mathbf{H}}$ are complex Gaussian random variables with zero mean and unit variance modelling the small-scale fading. Furthermore, we assume that the correlation coefficient between $\mathbf{h}_{k,c}$ and $\mathbf{h}_{k,e}$ is 0.9 for $k \in \mathcal{K}$. The correlation coefficient between $\mathbf{g}_{k,c}$ and $\mathbf{g}_{k,e}$ is $0.9$ for $k \in \mathcal{K}$. The correlation coefficient between $\mathbf{h}_{k_1,i_1}$ ($\mathbf{g}_{k_1,i_1}$) and $\mathbf{h}_{k_2,i_2}$ ($\mathbf{g}_{k_2,i_2}$) is $0$ for $k_1\neq k_2$, $i_1 \in \{c,e\}$, $i_2 \in \{c,e\}$. The noise power is $\sigma^2 = -80$ dBm. The target transmission rates of
$U_{k,c}$ and $U_{k,e}$ are $r_{k,c}= r_c$ and $r_{k,e}= r_e$, respectively, for $k \in \mathcal{K}$. The reflection coefficient vector $\bm{\phi}$ in \eqref{q10}, if not specified, is constructed using Case II. All simulation results are calculated based on an average of $100$ independent channel realizations and the convergence precision is $10^{-3}$.

In Fig. 1, we present the total transmit power at the BS, $P$, achieved by our proposed SOCP-ADMM based algorithm versus the number of iterations where $M=30$ and $N=8$. From Fig. 1, it is observed that, after about $6 \sim 8$ iterations, our proposed SOCP-ADMM based algorithm achieves the stable transmit power at the BS.

\begin{figure}
\centering
\includegraphics[width=3.6in]{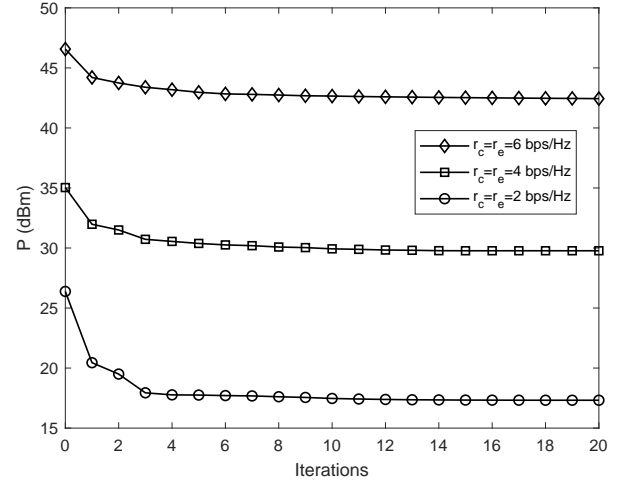}
\caption{Transmit power at the BS, $P$, versus the number of iterations; convergence performance of our proposed SOCP-ADMM based algorithm.}
\end{figure}
	
In Fig. 2, we present the transmit power comparison of different beamforming design schemes, where $r_c=r_e=4$ bps/Hz and $N = 8$. In the legend, ``SOCP-ADMM" denotes our proposed SOCP-ADMM based algorithm, ``ZF" denotes our proposed ZF based suboptimal algorithm, ``SDP" denotes the conventional SDP based algorithm proposed in \cite{QWu}, where $(\mathbf{w}_{k,c}, \mathbf{w}_{k,e})$ and $\bm{\phi}$ are alternatively optimized using the SDP, ``SDMA" denotes the conventional spatial division multiple access (SDMA) where a single user is served on each spatial direction \cite{ZDing03}, ``NOMA w/o IRS" and ``SDMA w/o IRS" denote the NOMA and SDMA schemes without IRS. From Fig. 2, it is found that our proposed SOCP-ADMM based algorithm outperforms the ``SDP" scheme. This is because in the ``SDP" scheme proposed in \cite{QWu}, the Gaussian randomization is required for rank-one recovery which degrades system performance. Furthermore, when $M$ is larger than 25, our proposed ZF-based suboptimal algorithm outperforms the ``SDP" scheme.
	
\begin{figure}
\centering
\includegraphics[width=3.6in]{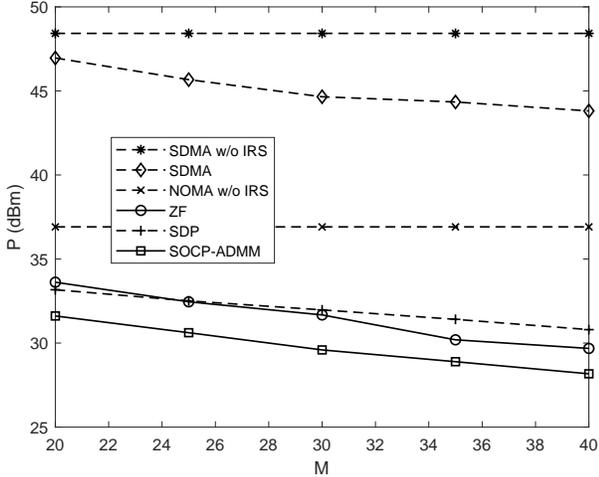}
\caption{Transmit power at the BS, $P$, versus the number of antennas at the IRS, $M$; performance comparison of different beamforming design schemes, where $r_c=r_e=4$ bps/Hz and $N = 8$.}
\end{figure}

In Fig. 3, we present the transmit power versus the number of antennas at the BS, $N$, where $r_c=r_e=4$ bps/Hz and $M =30$. From Fig. 3, it is shown that the transmit power at the BS decreases with the increase of $N$. Furthermore, the NOMA schemes, i.e., ``NOMA w/o IRS", ``ZF", ``SDP", and ``SOCP-ADMM" schemes, can
achieve much lower transmit power values compared with the SDMA schemes. This is expected since the NOMA schemes support more users to be served on each
spatial direction whereas the SDMA schemes serve only one user on
each spatial direction.

\begin{figure}
\centering
\includegraphics[width=3.6in]{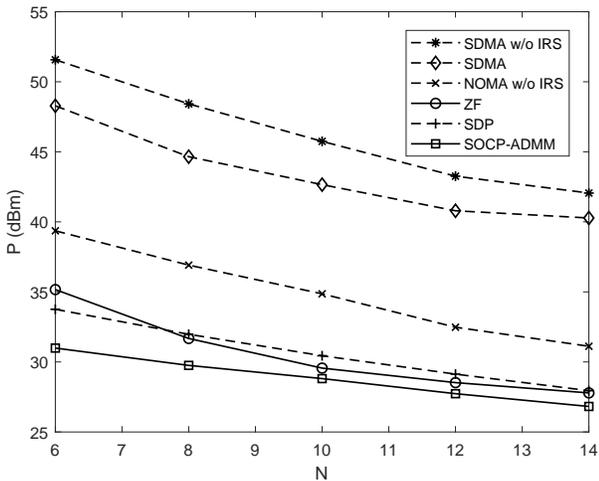}
\caption{Transmit power at the BS, $P$, versus the number of antennas at the BS, $N$; performance comparison of different beamforming design schemes, where $r_c=r_e=4$ bps/Hz and $M=30$.}
\end{figure}

In Fig. 4, we present the transmit power versus the target transmission rates of
central users, $r_c$, where $r_e=4$ bps/Hz, $M =30$, and $N=8$. From Fig. 4, it is observed that the transmit power at the BS increases as the transmission rate constraints at central users become more stringent. Furthermore, it is found from Fig. 4 that benefiting from the enhanced combined-channel strength from both the BS and the IRS, the IRS-aided NOMA can decrease the transmit power compared with the conventional NOMA scheme without IRS.

\begin{figure}
\centering
\includegraphics[width=3.6in]{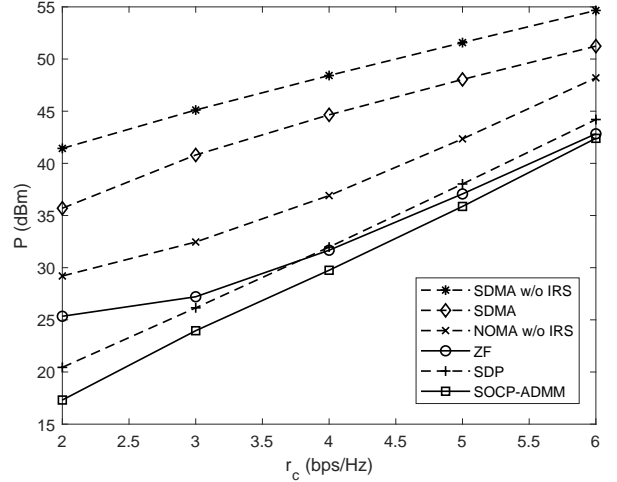}
\caption{Transmit power at the BS, $P$, versus the target transmission rates of
central users, $r_c$; performance comparison of different beamforming design schemes, where $r_e=4$ bps/Hz, $M = 30$ and $N=8$.}
\end{figure}

In Fig. 5, we present the transmit power at the BS versus the number of antennas at the IRS, $M$, when the reflection coefficient vector $\bm{\phi}$ is constructed based on Case I, Case II, and Case III, in \eqref{q10}, where $r_c=r_e=4$ bps/Hz and $N = 8$. For Case III, $L=2$, $L=4$, and $L=8$ are considered. From Fig. 5, it is illustrated that Case I achieves the minimum transmit power. This is because that Case I is the ideal case when $\bm{\phi}$ has infinite phase and amplitude resolution. For practical communication systems, the IRS has finite phase and amplitude resolution which generally degrades the system performance. Furthermore, it is observed from Fig. 5 that for Case III, as the phase resolution increases, i.e., $L$ increases, the performance gap between Case II and Case III becomes narrower.

\begin{figure}
\centering
\includegraphics[width=3.6in]{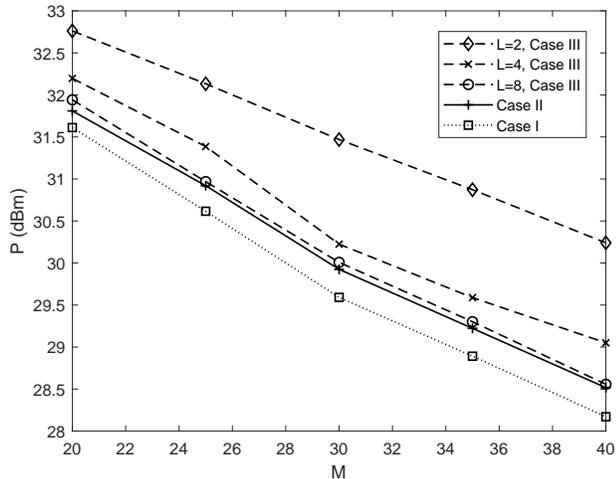}
\caption{Transmit power at the BS, $P$, versus the number of antennas at the IRS, $M$; performance comparisons when the reflection coefficient vector $\bm{\phi}$ is constructed is constructed based on Case I, Case II, and Case III, where $r_c=r_e=4$ bps/Hz and $N = 8$.}
\end{figure}

\section{Conclusion}

In this paper, we have proposed an effective SOCP-ADMM based algorithm for a multi-cluster MISO NOMA IRS-aided downlink communication network. To reduce the computational complexity, we have also proposed a low-complexity ZF based suboptimal algorithm. Simulation results show that our proposed SOCP-ADMM based algorithm outperforms the conventional SDP based algorithm. Furthermore, when the number of passive reflection elements is relatively high, our proposed ZF-based suboptimal algorithm also has superior system performance than the SDP based algorithm.

\end{document}